\documentclass[10pt,letterpaper]{article}
\usepackage[top=0.85in,left=2.75in,footskip=0.75in]{geometry}

\usepackage{changepage}

\usepackage[utf8]{inputenc}

\usepackage{textcomp,marvosym}

\usepackage{fixltx2e}

\usepackage{amsmath,amssymb}

\usepackage{cite}

\usepackage{nameref,hyperref}

\usepackage[right]{lineno}

\usepackage{microtype}
\DisableLigatures[f]{encoding = *, family = * }

\usepackage{rotating}

\usepackage{multirow,booktabs}
\usepackage{amsmath,amssymb,latexsym,MnSymbol}
\usepackage{hyperref}

\raggedright
\usepackage[aboveskip=1pt,labelfont=bf,labelsep=period,justification=raggedright,singlelinecheck=off]{caption}

\makeatletter
\renewcommand{\@biblabel}[1]{\quad#1.}
\makeatother

\date{}

\usepackage{lastpage,fancyhdr,graphicx}
\usepackage{epstopdf}
\fancyheadoffset[L]{2.25in}
\fancyfootoffset[L]{2.25in}
\begin{document}
\vspace*{0.35in}

\begin{flushleft}
{\Large
\textbf\newline{Rural to urban population density scaling of crime and property transactions in English and Welsh Parliamentary Constituencies}
}
\newline
\\
Quentin S. Hanley\textsuperscript{1,*},
Dan Lewis\textsuperscript{2},
Haroldo V. Ribeiro\textsuperscript{3},
\\
\bigskip
\bf{1} School of Science and Technology, Nottingham Trent University, Clifton Lane, Nottingham NG11 8NS, United Kingdom
\\
\bf{2} UkCrimeStats, Economic Policy Centre, London, SE1 3GA
\\
\bf{3} Departamento de F\'isica, Universidade Estadual de Maring\'a, Maring\'a, PR 87020-900, Brazil
\\
\bigskip

%
%





* \url{quentin.hanley@ntu.ac.uk}

\end{flushleft}
\section*{Abstract}
Urban population scaling of resource use, creativity metrics, and human behaviors has been widely studied. These studies have not looked in detail at the full range of human environments which represent a continuum from the most rural to heavily urban. We examined monthly police crime reports and property transaction values across all 573 Parliamentary Constituencies in England and Wales, finding that scaling models based on population density provided a far superior framework to traditional population scaling. We found four types of scaling: $i)$ non-urban scaling in which a single power law explained the relationship between the metrics and population density from the most rural to heavily urban environments, $ii)$ accelerated scaling in which high population density was associated with an increase in the power-law exponent, $iii)$ inhibited scaling where the urban environment resulted in a reduction in the power-law exponent but remained positive, and $iv)$ collapsed scaling where transition to the high density environment resulted in a negative scaling exponent. Urban scaling transitions, when observed, took place universally between 10 and 70 people per hectare. This study significantly refines our understanding of urban scaling, making clear that some of what has been previously ascribed to urban environments may simply be the high density portion of non-urban scaling. It also makes clear that some metrics undergo specific transitions in urban environments and these transitions can include negative scaling exponents indicative of collapse. This study gives promise of far more sophisticated scale adjusted metrics and indicates that studies of urban scaling represent a high density subsection of overall scaling relationships which continue into rural environments. 



\section*{Introduction}

Scaling in the evolution and development of cities has been widely studied~\cite{Pumain,Bettencourt,Samaniego,Arbesman,Bettencourt2,Bettencourt3,Mantovani,Gomez-Lievano,OliveiraCO2,Alves,Mantovani2,Alves2,Pan,Alves3,Ignazzi,Louf2,Melo,Rocha,BettencourtN1,Schlapfer,RybskiCO2,Masucci,vanRaan,Arcaute} with scaling behavior providing indicators of resource needs and productivity~\cite{Bettencourt3,Lobo,Alves4,Youn}. Cities promote innovation evidenced by super-linear scaling, while providing economies of scale in areas such as petrol stations and road surface~\cite{Bettencourt}. While super-linear scaling is beneficial in the production of new inventions, GDP and R \& D employment, cities also exhibit super-linear scaling of undesirable behaviors like homicide and violence~\cite{Bettencourt3,Alves,Alves2,Ignazzi}. Extensive study of human environments has led to the urban scaling hypothesis~\cite{Bettencourt4} which considers that some properties of cities change with size in scale invariant ways.  Although scaling behaviour follows similar mathematical forms, urban scaling parameters are not universal with coefficients varying between countries. For example, population scaling of homicide in cities in Colombia, Brazil, Mexico, and the United States vary widely~\cite{Bettencourt3,Alves,Alves2,Ignazzi} as well as vary over time for several urban metrics in Brazil~\cite{Alves4}. 

Region boundary definitions useful for understanding urban scaling remain challenging. Bettencourt proposed a combination of population and infrastructure leading to social co-ordination and efficiencies as the most important properties of cities~\cite{Bettencourt5}. However, these key features do not correspond to traditional definitions of cities which have more provincial origins. For example, formal City status in the UK is given by the Monarch as an honorary designation~\cite{CabinetOffice}. Urban administrative units consist of a mix of Unitary Authorities, Metropolitan Regions, Boroughs, and other units like the Greater London Authority. Within these designations, there may be several ``cities'' in a contiguous urban area. As a result, a robust definition of cities and urban regions is needed in order to fully test the urban scaling hypothesis.  In the UK context, a model of cities was recently proposed based on population density thresholds and commuter flow rather than total population~\cite{Arcaute}. This challenges  the concept of population as the key predictor of urban scaling~\cite{Masucci,Arcaute}. Elsewhere, related studies~\cite{Cottineau} indicate scaling exponents are sensitive to whether urban areas are defined by built up areas or built up areas together with a surrounding commuter zone.  Even with such definitions of cities and urban areas, the resulting regions vary in their characteristics. The Tokyo-Yokohama region of Japan with a population in excess of 30 million has a density of 44 people per hectare (p/ha) while Dhaka in Bangladesh has fewer people but a population density of 435 p/ha~\cite{Cox}. Similarly, urban areas are not uniform across their footprint. They include regions with very low resident populations such as parks and industrial areas.

Cities are important human environments and the proportion of the world's population living in urban areas has been consistently increasing. As of 2014, the urban population reached 54\% of the total; however, the number of people residing outside of cities is large and countries range from 100\% (Singapore) to 8.6\% (Trinidad and Tobago) urban~\cite{UN}. Many studies of urban environments exist, however, nearly universally, these implicitly assume a fundamental difference between cities and surrounding regions while not examining more rural environments in detail. In England and Wales, urban areas are connected built-up areas containing at least 10,000 people~\cite{Bibby}. All other areas are considered to be rural. Notably, England in 2011 had  82.4\% of people living in urban areas. However, rural areas made up 85\% of the land~\cite{GovtStatServ}. This leaves an open question related to the more general applicability of scaling laws obtained from cities to understand the full range of human environments.  

Crime is known to follow scaling laws in urban areas. The self-similarity of cities underlying the urban scaling hypothesis is unlikely to change fundamentally based on the outcome of current debates about the definition of cities. The acceleration of crime described by power-law scaling will probably remain; however, a broader understanding of crime scaling over a greater range of human environments can provide great insight into scaling phenomena of all types.  

The UK government has published extensive data on property transactions and police reported crime. These data are notable for their high spatial accuracy and good coverage within England and Wales. Although police reported crime has been the subject of considerable controversy in the UK~\cite{Patrick,Anonymous}, the data set is extensive and can be mapped into a range of shapes. This provides a unique opportunity to interrogate the scaling of crime reports and property transactions over the full surface area of England and Wales.

Here we investigate scaling relationships for a range of crime types and property transactions values in England and Wales broken down by Parliamentary Constituencies.  Parliamentary Constituencies cover a wide range of communities from the most rural to heavily urban and are well defined. This allows scaling to be studied using a wide range of crime and property metrics. Using a continuum of rural and urban environments, the extent to which cities extracted from their surroundings are sufficient to understand the scaling of human economic and criminal behaviors can be assessed.   

\section*{Methods}
\subsection*{Data Sets}
We have accessed data at the Parliamentary Constituencies level of all 573 constituencies in England and Wales (see next section). These data are composed of population $N$, daytime population $N_d$, constituency area $A$, 15 crime types and transaction value of 9 property types (Table~\ref{tab:0}). The population data were obtained from the Ordnance Survey mid-2013 estimates, and daytime population estimates from the Office of National Statistics (\url{www.ons.gov.uk}). The constituency boundary areas were calculated from geographic shape files of the Ordnance Survey Boundary Line dataset. Crime data were obtained from the Home Office via their open data portal (\url{https://data.police.uk/}). Property data were obtained from the Land Registry. These data were collated on the UKCrimeStats (\url{http://www.ukcrimestats.com/}) data platform and provided as monthly reports. The crime data from 2014 were captured on 10/6/2015 and property transaction value data on 17/7/2015. Prior to analysis the monthly values from each constituency were summed over the 12 months of study. If a constituency did not have any crime or property transaction of a particular type over the 12 month period it was removed from the analysis. Only the  Cities of London and Westminster (Semi-detached) and Bethnal Green and Bow (Detached) in England reported no property transactions of a particular type in the period and were dropped from the respective property analyses. The entire data set is maintained and made freely available by the Office of National Statistics and the UK Home Office. As these data are subject to updates, the snapshot has been provided as \nameref{S1_dataset}.

\begin{table}[!ht]

\caption{\textbf{Crime and property types analyzed in this study.}}
\begin{adjustwidth}{1.5cm}{0in}
\begin{tabular}{l|l}
\hline
\multicolumn{2}{c}{Constituency metrics, $Y$}\\
\hline
{Crime type} & {Property Type}\\
\hline
Anti-Social Behavior (ASB) & Detached\\
Bike Theft & Flats\\
Burglary & Freehold\\
Criminal Damage and Arson (CD \& A)  & Leasehold \\
Drugs & New \\
Order & Old\\
Other Crime & Semi-detached\\
Other Theft & Terraced \\
Robbery & Total Property \\
Shoplifting & \\
Theft from the Person & \\
Total Crime and ASB & \\
Vehicle Crime & \\
Violence & \\
Weapons & \\
\hline
\end{tabular}
\label{tab:0}
\end{adjustwidth}
\end{table}

\subsection*{Overview of Parliamentary Regions}
Parliamentary Constituencies were selected as regions with clearly defined shapes and similar populations while not being exclusively urban. Parliamentary constituency data were obtained for all 573 constituencies in England and Wales. The regions ranged in area from 331,440 ha (Penrith and The Border) down to 738 ha (Islington North). Constituency populations were from 56,651 (Aberconwy, Wales) to 163,398 (West Ham, England) while population density ranged from 0.22 people per hectare (Brecon and Radnorshire, Wales) up to 150 p/ha (Westminster North, England). Similar values for daytime population were from 55,453 (Aberconwy, Wales) to 946,397 (Cities of London and Westminster, England) and daytime population densities from 0.22 p/ha (Brecon and Radnorshire, Wales) to 550.3 p/ha (Cities of London and Westminster, England).  This range of population densities includes regions that exceed the density of many of the world's largest cities when considered as a whole. It is notable that constituency populations for England and Wales fall within a factor of 3; however, total reported crime and anti-social behavior varied by a factor of 17 and total property transactions by a factor of 65.

\section*{Results and Discussion}
Urban power-law scaling has been observed in many parts of the world~\cite{Pumain,Bettencourt,Samaniego,Arbesman,Bettencourt2,Bettencourt3,Mantovani,Gomez-Lievano,OliveiraCO2,Alves,Mantovani2,Alves2,Pan,Alves3,Ignazzi,Louf2,Melo,Rocha,BettencourtN1,Schlapfer,RybskiCO2,Masucci,vanRaan}. Aspects remain controversial in part due to uncertainty about how best to define cities and concern about the use of population as a definitive metric~\cite{Arcaute}. Bettencourt \textit{et al.}~\cite{Bettencourt} defined the urban scaling of a particular metric at a particular time as
\begin{equation}\label{eq_usualscaling}
Y= Y_0\, N^\beta~~{\text{or its linearized version}~~} \log Y = \log Y_0 + \beta \log N\,.
\end{equation}
In this, $Y$ is a metric (\textit{e.g.} energy, patents, serious crime), $Y_0$ is a constant, $N$ is the population, and $\beta$ the power-law (or allometric) exponent. When $\beta < 1$ the metric decreases proportionally with scale (such as road surface or petrol stations) and when $\beta > 1$ the metric accelerates (examples include GDP and new AIDS cases). 

\begin{figure}[!ht]
\begin{adjustwidth}{-2.25in}{0in}
\begin{center}
\includegraphics[scale=0.35]{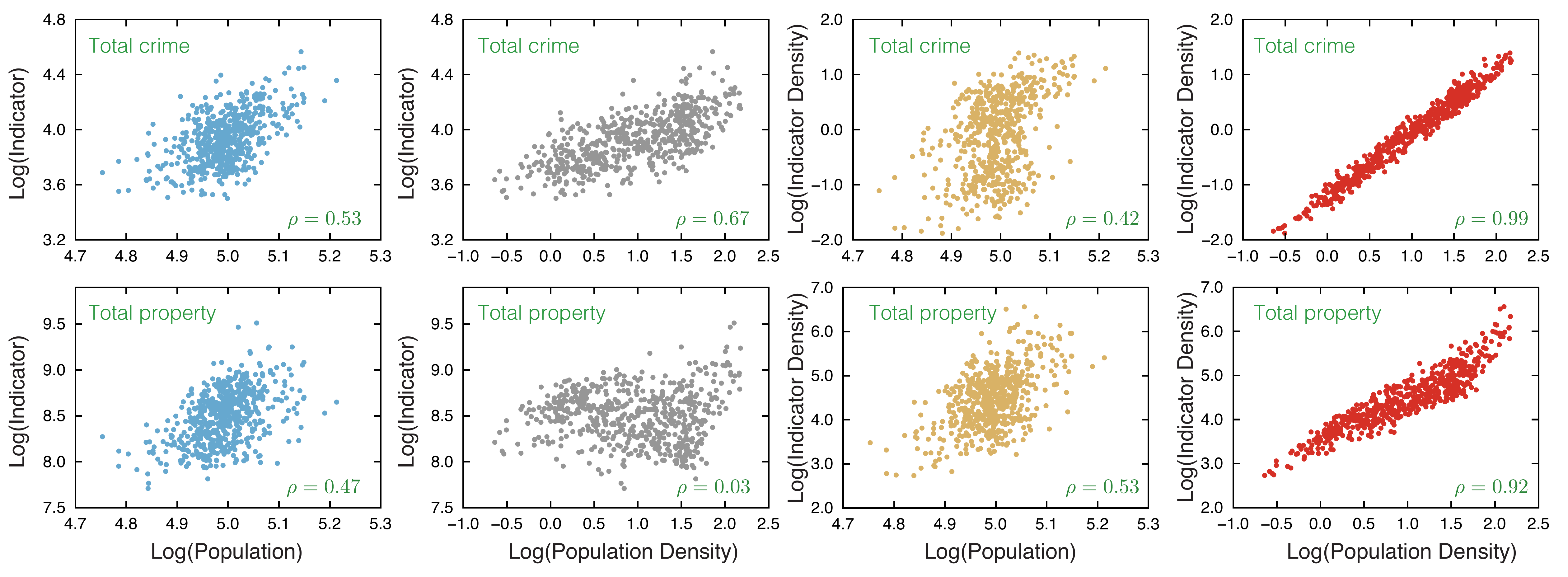}
\end{center}
\caption{
\textbf{Comparison of predictor and indicator metrics for the indicators total crime and total property transaction values.} The density metrics gave better correspondence to the scaling laws as indicated by Pearson correlation ($\rho$). }
\label{fig:1}
\end{adjustwidth}
\end{figure}

The form of Eq.~\ref{eq_usualscaling} can be adapted to consider other metrics. For our data, the scaling behavior of property transaction values and police reported crime were tested by comparing 8 models considering population, daytime population, population density, and daytime population density as predictors of crime and property metrics expressed directly (\textit{e.g.} number of crimes) or as a density (crimes per hectare). For instance, when considering both population and indicator density, Eq.~\ref{eq_usualscaling} can be rewritten as 
\begin{equation}\label{eq_usualscaling_d}
\log y = \log y_0 + \beta \log d\,,
\end{equation}
where $y=Y/A$ is the indicator density (\textit{e.g.} a particular crime per hectare) and $d=N/A$ is the population density. Figure~\ref{fig:1} illustrates some of these models by showing scatter plots of $\log Y \times \log N$, $\log Y \times \log d$, $\log y \times \log N$ and $\log y \times \log d$ for the metrics total crime and total property value. By including all categories of crime and property together in a single analysis, we found that the density metrics were superior with daytime population density slightly better for prediction of crime and resident population density better for predicting property transaction values. For this set of metrics, both $R^2$ and predicted residual sum of squares (PRESS) statistics from general prediction models confirmed the density metrics were superior (Table~\ref{tab:1}). 

\begin{table}[!ht]
\begin{adjustwidth}{-2.25in}{0in}
\caption{\textbf{Comparison of metrics for prediction of crime and property transaction values.} All models included categorical variables describing the type of crime or property as: Predictor, Type, Predictor*Type. The model with the best $R^2$ and PRESS statistics have been highlighted in bold. }
\centering
\begin{tabular}{llrr}
\hline
{Dependent } & {Predictor} & $R^2$ (\%)  & PRESS \\
\hline
Log(Crime) & Log(Population) & 84.97 & 541\\
Log(Crime) & Log(Daytime Population) & 85.70 & 519 \\
Log(Crime Density) & Log(Population Density) & 95.36 & 380 \\
\textbf{Log(Crime Density)} & \textbf{Log(Daytime Population Density)} & \textbf{95.92} & \textbf{333} \\
Log(Transaction Value) & Log(Population) & 59.63 & 703 \\
Log(Transaction Value) & Log(Daytime Population) & 55.44 & 774 \\
\textbf{Log(Transaction Value Density)} & \textbf{Log(Population Density)} & \textbf{80.31} & \textbf{689} \\
Log(Transaction Value Density) & Log(Daytime Population Density) & 79.90 & 703 \\
\hline
\end{tabular}
\label{tab:1}
\end{adjustwidth}
\end{table}

\begin{figure}[!ht]
\begin{adjustwidth}{-2.25in}{0in}
\begin{center}
\includegraphics[scale=0.32]{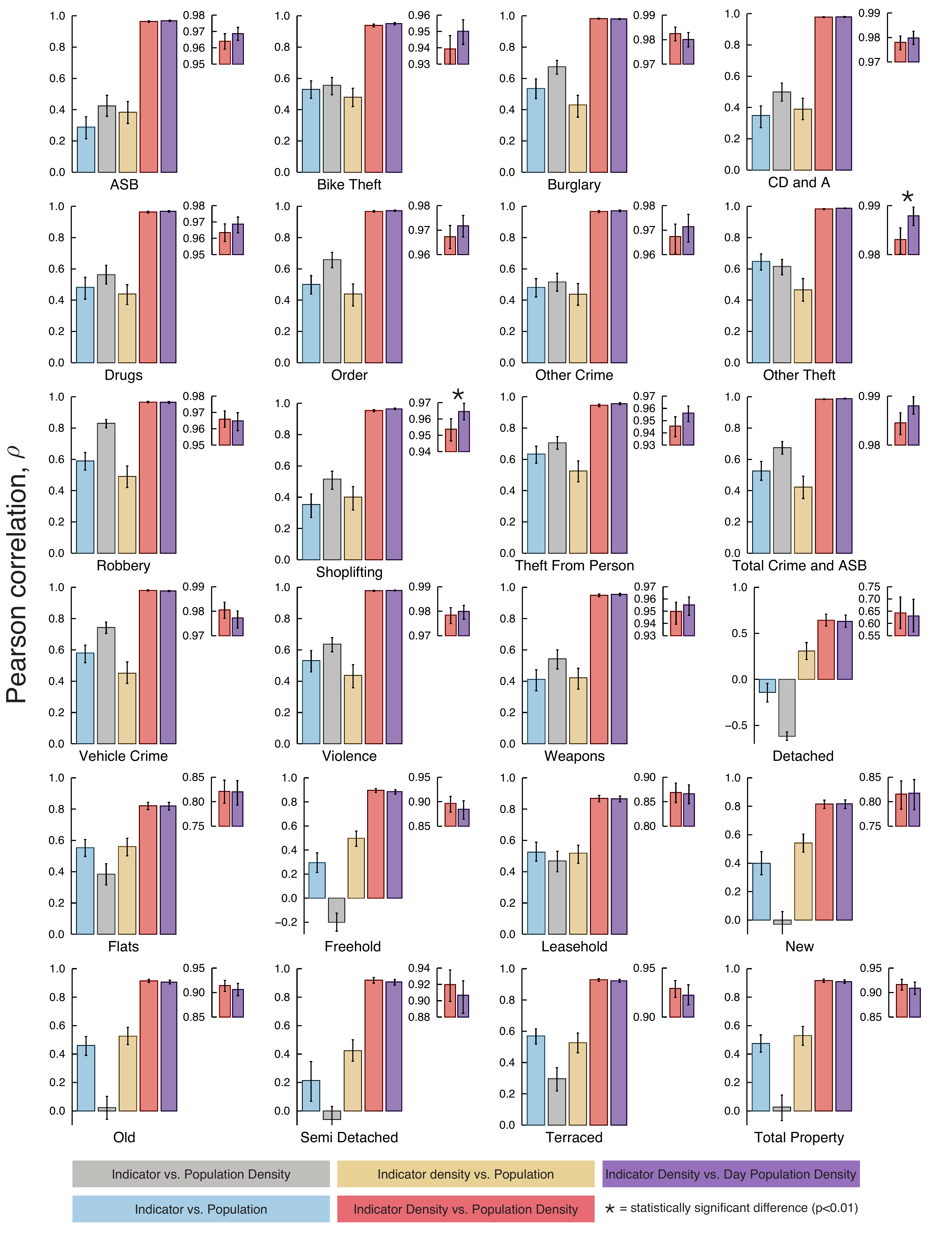}
\end{center}
\caption{\textbf{Comparison of Pearson correlations for different property and crime types.} Markedly improved correlations are observed using density metrics which were superior in all cases. Here the error bars stand for 99\% confidence interval obtained via bootstrap and the asterisk marks indicate a significative difference between population density and day population density (via bootstrap two-sample mean test with 99\% confidence).}
\label{fig:2}
\end{adjustwidth}
\end{figure}

To appreciate the superiority of the density metrics, it is helpful to view the correlations in isolation (Fig~\ref{fig:2} and \hyperref[S1_Fig]{S1~Fig}). A large improvement in correlation is seen when moving to density metrics and in some cases this changed the sign of the correlation. This was also apparent in the general models where a qualitative change from models dominated by categorical variables to ones dominated by continuous variables was observed when density was used. The improvement obtained from population density metrics was not surprising given the data set used. Parliamentary Constituencies were chosen due to having relatively small variations in total population while varying greatly in area. 

The shift from population density to daytime population density gave a comparatively marginal change in outcome (Fig~\ref{fig:2}). Across all property and crime categories, 13/24 were more highly correlated with daytime property density than resident population density, roughly the expectation if the two predictors were equal. However, property and crime had distinct profiles. For property, resident population density was always more highly correlated than daytime population density giving $p = 0.0078$ for a binomial test; however, it is not significantly better when considered in isolation (Fig~\ref{fig:2}). For crime, 12 out of 15 categories ($p = 0.0352$ for a binomial test) were more highly correlated with daytime population density with only 2 cases (Other Theft and Shoplifting) significant when considered in isolation. Other theft includes a range of non-violent theft offenses where large daytime crowds may facilitate commission of the crime. Also, we find no significant difference between population density and daytime population density for all property and crime categories when  considering the maximal information coefficient (MIC, \hyperref[S3_Fig]{S1~Fig})~\cite{Reshef}. As the improvement overall going to daytime population data was marginal and the availability of similar data across the world is limited, we focused on resident population density metrics in our subsequent presentation. 

As in the case shown for total property (Fig~\ref{fig:1}), we found that several density metrics displayed a more complex scaling behavior and a single power law (Eq.~\ref{eq_usualscaling}) was insufficient to describe the observed data.  Complex scaling has been observed in other types of scaling. For example, it has been noted in fluctuation scaling of crime~\cite{Hanley}, disease~\cite{Keeling}, and a variety of physical processes~\cite{Eisler} and scientists have been encouraged to test alternative models to power laws when appropriate~\cite{Clauset}. Here, visually inspired by the behavior of our data, we tested whether a double power-law provided a significantly better fit between a density metric ($y$) and the population density ($d$) than a single power law, that is,
\begin{equation}\label{eq_doublescaling}
\log y = 
\begin{cases}

\log y_0 + \beta_{\text{L}}\log d & (\text{for}~\log d\leq \log d^*)\\
\log y_0 + (\beta_{\text{L}}-\beta_{\text{H}})\log d & (\text{for}~\log d> \log d^*)\\
\end{cases}\,,
\end{equation}
where $d^*$ is a population density threshold, $\beta_{\text{L}}$ ($\beta_{\text{H}}$) is the power-law exponent for low (high) population density, $y_0$ and $y_1$ are constants. In particular, we have chosen $\log y_1 = \log y_0 + (\beta_{\text{L}}+\beta_{\text{H}})\log d^*$, holding the continuity of $y(d)$. Thus, the model of Eq.~\ref{eq_doublescaling} has two additional parameters when compared with the single power-law model of Eq.~\ref{eq_usualscaling_d}. This approach provides a picture of the data based on the prevailing view of population scaling (a single power law) against a simple alternative of a double power law.  In all cases, the parameters reported were highly significant, which does not rule out that another function or set of functions may fit the data better.

\begin{figure}[!ht]
\begin{adjustwidth}{-2.25in}{0in}
\begin{center}
\includegraphics[scale=0.3]{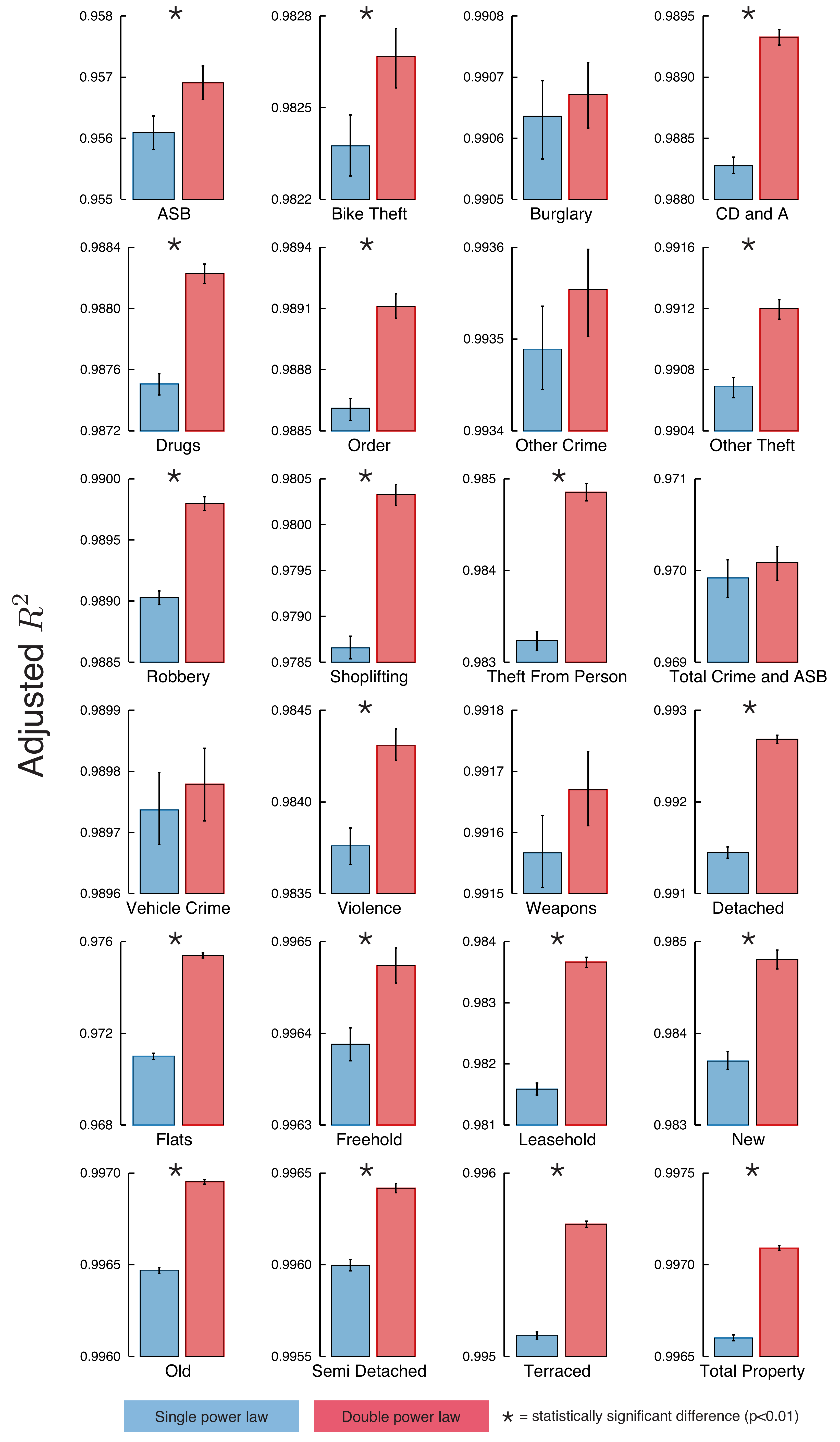}
\end{center}
\caption{\textbf{Comparison of the adjusted $R^2$ obtained for the single power-law model (Eq.~\ref{eq_usualscaling_d}) and the double power-law model (Eq.~\ref{eq_doublescaling}).} Error bars stand for 99\% bootstrap confidence intervals and the asterisk marks indicate a significant difference (via bootstrap two-sample mean test with 99\% confidence). Notice that the double power-law model is a better fit in 19 out 24 metrics; however, for other crime, total crime and ASB, vehicle crime, and weapons the differences in the adjusted $R^2$ are not statistically significant. 
}
\label{fig:3}
\end{adjustwidth}
\end{figure}

We compared the models provided by Eqs.~\ref{eq_usualscaling_d} and~\ref{eq_doublescaling} and tested whether the double power-law model gave statistically significant improvement. For the single power-law (Eq.~\ref{eq_usualscaling_d}), we employed ordinary least squares regression in the log transformed data for obtaining the parameters $y_0$ and $\beta$ as well as the adjusted $R^2$. We then used bootstrapping to determine the confidence intervals for the adjusted $R^2$. Simulated annealing~\cite{Kirkpatrick} was used for fitting the double power-law model (Eq.~\ref{eq_doublescaling}) to the log transformed data by considering the residual sum of squares as the cost function, yielding the parameters $y_0$, $y_1$, $\beta_{\text{L}}$ and $\beta_{\text{H}}$, and also the adjusted $R^2$. Again, the confidence intervals for the adjusted $R^2$ were calculated via bootstrapping. We further considered two-sample bootstrap tests for testing the null hypothesis that the adjusted $R^2$ from Eqs.~\ref{eq_usualscaling_d} and~\ref{eq_doublescaling} are equal~\cite{Efron}. Figure~\ref{fig:3} compares the values of the adjusted $R^2$ for both models, where we noticed that double power-law model is superior in 19 out 24 metrics. Similar conclusions regarding the model selection were obtained by considering the Akaike Information Criterion (AIC) or Bayesian Information Criterion (BIC)~\cite{Burnham} (\hyperref[S2_Fig]{S2} and \hyperref[S3_Fig]{S3~Figs}). 

\begin{figure}[!ht]
\begin{adjustwidth}{-2.25in}{0in}
\begin{center}
\includegraphics[scale=0.33]{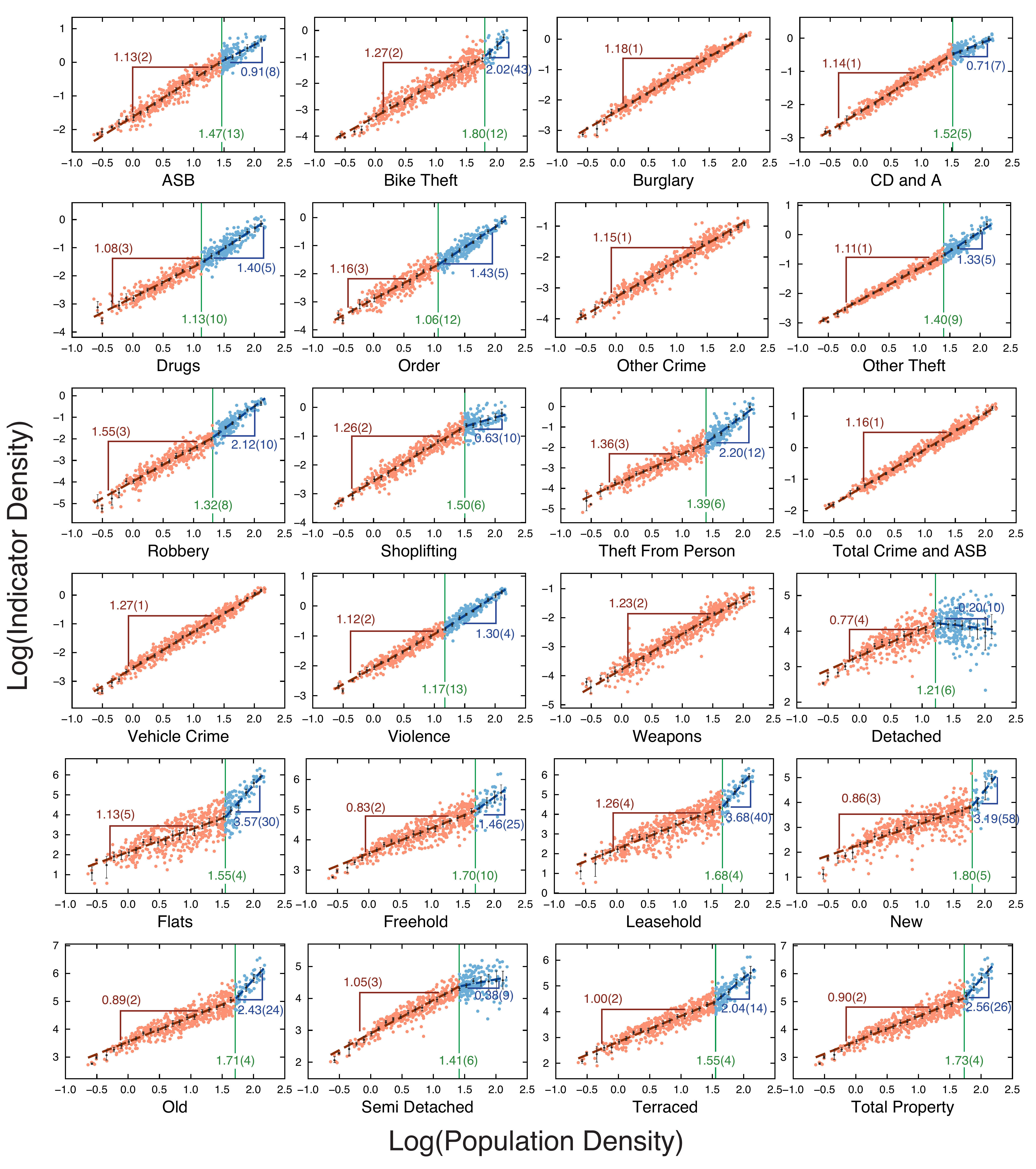}
\end{center}
\caption{\textbf{Population density scaling behavior of all metrics.} The colorful dots are the empirical values and the black dots are the window average values (errors bars are 95\% bootstrap confidence intervals). For metrics in which the double power-law is a better fit according to adjusted $R^2$ (see also \hyperref[S4_Fig]{S4~Fig} for AIC and BIC), the red dots show the low-density data ($\log d< \log d^*$) and blue ones the high density data. The vertical line represents the position ($\log d^*$) of the transition to urban scaling. The numbers refer to the scaling law exponents and the threshold positions with uncertainty in parentheses. Total crime reports were 5,357,113. Total property transactions were £226,122,179,000.}
\label{fig:4}
\end{adjustwidth}
\end{figure}

The improvement in the scaling laws using the density metrics thus revealed segmented scaling in several but not all metrics (Fig~\ref{fig:4}) indicating the onset of complex scaling (Eq.~\ref{eq_doublescaling}). The scaling parameters are shown in Table~\ref{tab:2}, where we observe that 5 crime metrics followed a single scaling law over all densities with no evidence for a specifically ``urban'' scaling law, only a continuation of low density behavior. The remaining metrics all exhibited complex scaling and all thresholds fell between 10 and 70 p/ha. 

\begin{table}[!ht]
\begin{adjustwidth}{-2.25in}{0in}
\caption{\textbf{Scaling parameters for police crime report density and property transaction value density with population density.}}
\centering
\begin{tabular}{lrrrrr}
\hline
{Crime Type} & \multicolumn{1}{c}{$\log(y_0)$} & \multicolumn{1}{c}{$\beta_{\text{L}}$ or $\beta$} & \multicolumn{1}{c}{$\log(y_1)$} & \multicolumn{1}{c}{$\log(d^*)$} & \multicolumn{1}{c}{$\beta_{\text{H}}$}\\
\hline
ASB & $-1.62\pm0.02$ & $1.13\pm0.02$ & $-1.30\pm0.13$ & $1.47\pm0.13$ & $0.91\pm0.08$\\
Bike Theft & $-3.26\pm0.02$ & $1.27\pm0.02$ & $-4.62\pm0.77$ & $1.80\pm0.12$ & $2.03\pm0.43$\\
Burglary & $-2.35\pm0.01$ & $1.18\pm0.01$ & \multicolumn{1}{c}{-} & \multicolumn{1}{c}{-} & \multicolumn{1}{c}{-}\\
CD and A & $-2.21\pm0.01$ & $1.14\pm0.01$ & $-1.55\pm0.11$ & $1.52\pm0.05$ & $0.71\pm0.07$\\
Drugs & $-2.77\pm0.02$ & $1.08\pm0.03$ & $-3.13\pm0.08$ & $1.13\pm0.10$ & $1.40\pm0.05$\\
Order & $-2.91\pm0.02$ & $1.16\pm0.03$ & $-3.20\pm0.07$ & $1.06\pm0.12$ & $1.43\pm0.05$\\
Other Crime & $-3.29\pm0.01$ & $1.15\pm0.01$ & \multicolumn{1}{c}{-} & \multicolumn{1}{c}{-} & \multicolumn{1}{c}{-}\\
Other Theft & $-2.26\pm0.01$ & $1.11\pm0.01$ & $-2.57\pm0.08$ & $1.40\pm0.09$ & $1.33\pm0.05$\\
Robbery & $-3.98\pm0.02$ & $1.55\pm0.03$ & $-4.73\pm0.14$ & $1.32\pm0.08$ & $2.12\pm0.10$\\
Shoplifting & $-2.56\pm0.02$ & $1.26\pm0.02$ & $-1.61\pm0.16$ & $1.50\pm0.06$ & $0.63\pm0.10$\\
Theft from the Person & $-3.68\pm0.03$ & $1.36\pm0.03$ & $-4.84\pm0.18$ & $1.39\pm0.06$ & $2.20\pm0.12$\\
Total Crime and ASB & $-1.22\pm0.01$ & $1.16\pm0.01$ & \multicolumn{1}{c}{-} & \multicolumn{1}{c}{-} & \multicolumn{1}{c}{-}\\
Vehicle Crime & $-2.54\pm0.01$ & $1.27\pm0.01$ & \multicolumn{1}{c}{-} & \multicolumn{1}{c}{-} & \multicolumn{1}{c}{-}\\
Violence & $-2.06\pm0.01$ & $1.12\pm0.02$ & $-2.28\pm0.06$ & $1.17\pm0.13$ & $1.30\pm0.04$\\
Weapons & $-3.78\pm0.02$ & $1.23\pm0.02$ & \multicolumn{1}{c}{-} & \multicolumn{1}{c}{-} & \multicolumn{1}{c}{-} \\
\hline
Property Type & & & & &\\
\hline
Detached & $3.30\pm0.03$ & $0.77\pm0.04$ & $4.47\pm0.14$ & $1.21\pm0.06$ & $-0.20\pm0.10$\\
Flats & $2.13\pm0.05$ & $1.13\pm0.05$ & $-1.65\pm0.48$ & $1.55\pm0.04$ & $3.57\pm0.30$\\
Freehold & $3.55\pm0.02$ & $0.83\pm0.02$ & $2.48\pm0.42$ & $1.70\pm0.10$ & $1.46\pm0.25$\\
Leasehold & $2.24\pm0.04$ & $1.26\pm0.04$ & $-1.83\pm0.69$ & $1.68\pm0.04$ & $3.68\pm0.40$\\
New & $2.30\pm0.03$ & $0.86\pm0.03$ & $-1.88\pm1.06$ & $1.80\pm0.05$ & $3.19\pm0.58$\\
Old & $3.55\pm0.02$ & $0.89\pm0.02$ & $0.92\pm0.42$ & $1.71\pm0.04$ & $2.43\pm0.24$\\
Semi Detached & $2.90\pm0.02$ & $1.05\pm0.03$ & $3.84\pm0.14$ & $1.41\pm0.06$ & $0.38\pm0.09$\\
Terraced & $2.83\pm0.02$ & $1.00\pm0.02$ & $1.23\pm0.22$ & $1.55\pm0.04$ & $2.04\pm0.14$\\
Total Property & $3.57\pm0.02$ & $0.90\pm0.02$ & $0.69\pm0.46$ & $1.73\pm0.04$ & $2.56\pm0.26$\\
\hline
\end{tabular}
\label{tab:2}
\end{adjustwidth}
\end{table}

\begin{figure}[!ht]
\begin{center}
\includegraphics[scale=0.3]{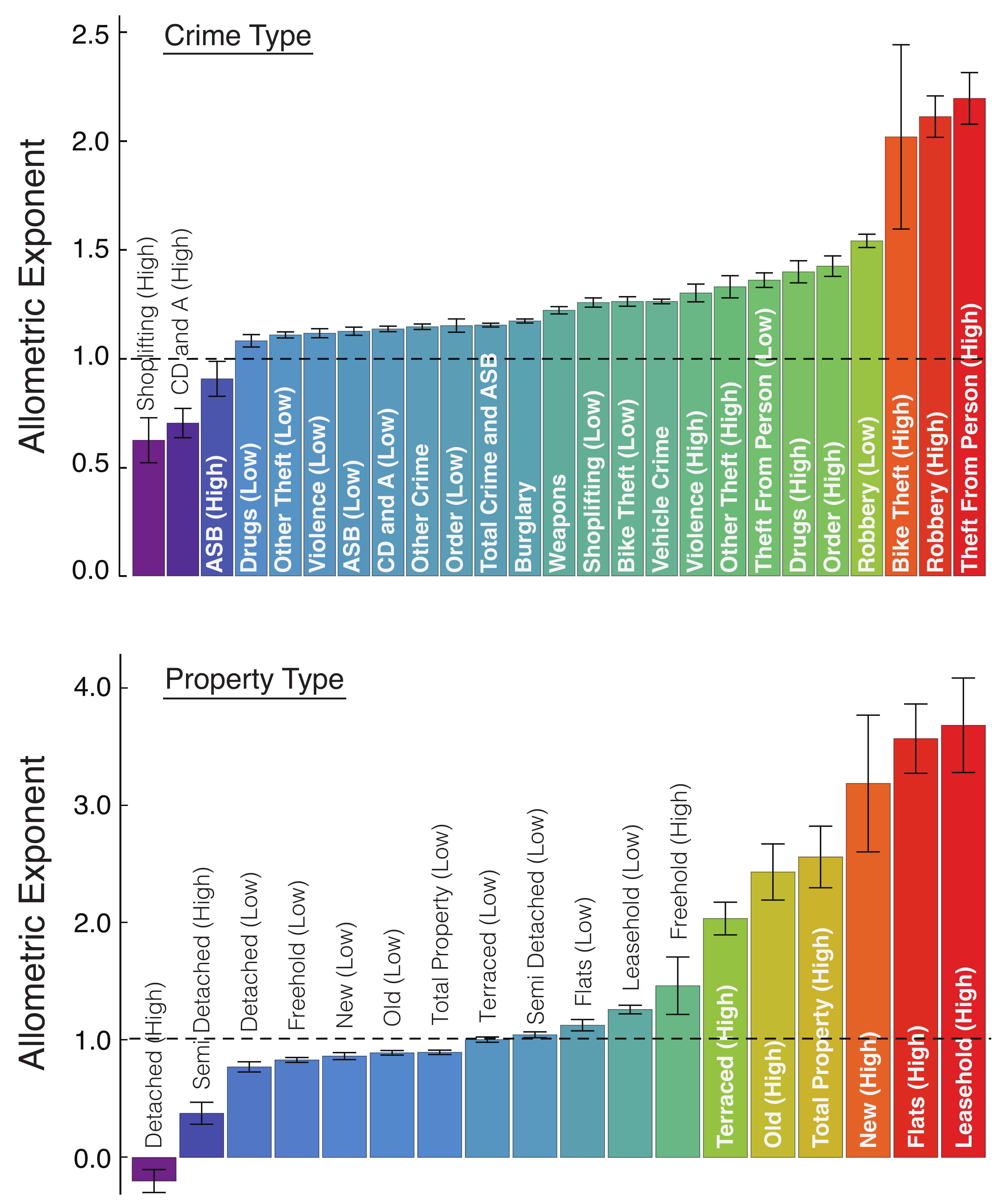}
\end{center}
\caption{\textbf{Allometric exponents for crime metrics (upper panel) property transactions (bottom panel) using density metrics.} The error bars refer to the standard errors in the exponents.}
\label{fig:5}
\end{figure}

Comparison of exponents (Figs~\ref{fig:4} and~\ref{fig:5}) revealed four types of density scaling including three  specifically related to ``urban effects'' . The ``non-urban scaling'' was found for  burglary, other crime, total crime and antisocial behavior, vehicle crime, and weapons. This designation was applied to metrics where no threshold value could be discerned in the data. Of the three types of urban scaling, the first is ``accelerated urban scaling'' where $\beta_{\text{L}} < \beta_{\text{H}}$. This was observed in the majority of metrics and applied to: bike theft, drugs, order, other theft, robbery, theft from the person, violence, flats, freehold, leasehold, new, old, terraced, and total property. Metrics following accelerated urban scaling are specifically enhanced in an urban environment. The second urban category is ``inhibited urban scaling'' ($\beta_{\text{L}}>\beta_{\text{H}} > 0$). Inhibited urban scaling was observed for anti-social behavior, criminal damage and arson (CD and A), shoplifting, and semi-detached properties. Metrics following inhibited urban scaling undergo specifically urban economies of scale. The last urban category is ``collapsed urban scaling'' ($\beta_{\text{L}}>\beta_{\text{H}}$, with $\beta_{\text{H}}<0$). Only a single category (detached housing) followed this type of scaling and to our knowledge this is the first time a negative exponent has been reported in the context of urban scaling. 

Scaling studies of many of the crime metrics used here have not  been reported nor have their transitions in urban environments. The variable effects of high population density are noteworthy. For example, criminal damage which undergoes inhibited urban scaling has been associated with binge drinking in the UK~\cite{Richardson} while property crimes including criminal damage have been linked to foreclosures in the US~\cite{Lacoe}.  Finding general scaling laws for such behaviour suggests many of these have a wider context. In the case of criminal damage and arson, opportunities appear to be reduced at high population density and high amounts of property crime associated with foreclosures may be a symptom of loss of an inhibitory population density rather than foreclosures directly. A detailed review aligning the scaling laws reported here with the extensive criminological literature should provide considerable insight. 

Specifically urban scaling phenomena were associated with transitions between 10 and 70 p/ha. The high end of the density thresholds (63 p/ha) exceeds the highest density threshold considered by Arcuate et al. (40 p/ha)~\cite{Arcaute}. It also exceeds the average population density of London (59 p/ha) and all the large cities in Europe and North America outside of Mexico (\textit{e.g.} Moscow (35 p/ha), Paris (38 p/ha), and Zurich (32 p/ha))~\cite{Cox}. It is noteworthy that the highest population density found in a US or Canadian city (Los Angeles) is 24 p/ha when considered as a whole~\cite{Cox}. This suggests that most of the transitions seen here may be unobservable in much of Europe and North America unless cities are subdivided into high density regions as was done here. 

\section*{Conclusion}
This study significantly refines the urban scaling hypothesis. It set out to investigate regions that are reasonably well matched in population to accentuate scaling behaviours that might arise from inhomogeneity within cities and other density related features. Despite relatively small population variation, there is support for the existing view of population scaling, however in this data set density metrics were universally better. For some metrics, a single power law is sufficient to explain scaling at all population densities over a continuum from rural to urban. These metrics are subject to a single rural-urban scaling law and, in such cases, the scaling behavior of human environments is simpler than previously thought. As there is no clear distinction to be made between urban and rural environments for these metrics, there is less need to define city boundaries precisely. For other metrics, there is indisputable evidence for specifically rural and specifically urban scaling.

The results indicate that many metrics are not scale invariant in what are currently understood as urban settings. Observed transitions from rural to urban behaviour were in the range of 10-60 p/ha which is roughly in the midrange of the top 1000 cities with $>$500,000 of population when sorted by population density~\cite{Cox}. These scaling transitions are associated with acceleration, inhibition, or collapse of the scaling law within the high population density environments of cities. Such behavior is intuitive for some metrics. For example, detached housing is clearly an unsustainable property type at high population density and a collapse in transactions of this type is unsurprising in a high density urban environment. Finding a transition at urban population densities clearly supports the notion of uniquely urban behavior underlying the urban scaling hypothesis. However, most currently published studies have not examined the low side of these density thresholds in detail and will miss the transition from rural to urban scaling. It is also of interest to do more extensive studies on the great cities of Asia, Africa, and the Americas south of the Mexico-USA border. Cities in these parts of the world have particularly high population densities not found in Europe and other parts of North America and may yield more interesting behavior. 

Implicit in the design of this study is the notion that both rural and urban environments are non-uniform. A city the size of London is heterogeneous in its distribution of population, property and crime. Greater London includes 73 constituencies allowing the non-uniformity of this region to be considered in the scaling models rather than as a single monolithic conurbation or metropolitan region.

Finally, this study adds evidence to the long-standing challenge to crime rates and per capita comparisons~\cite{Bettencourt3,Alves2,Ignazzi,Alves4}. It is clear that high or low per capita crime rates are uninterpretable outside of the context of the scaling law to which they belong and, based on the current study, similar considerations are appropriate for the study of property transactions.

\section*{Supporting Information}

\subsection*{S1 Dataset}
\label{S1_dataset}
{\bf Data employed in this study.} Snapshot of police reported crime captured 10/6/2015 and property transaction values captured  17/7/2015 for the 12 months of 2014. (XLSX)
\clearpage
\subsection*{S1 Fig.}
\label{S1_Fig}
\begin{center}
\includegraphics[scale=0.3]{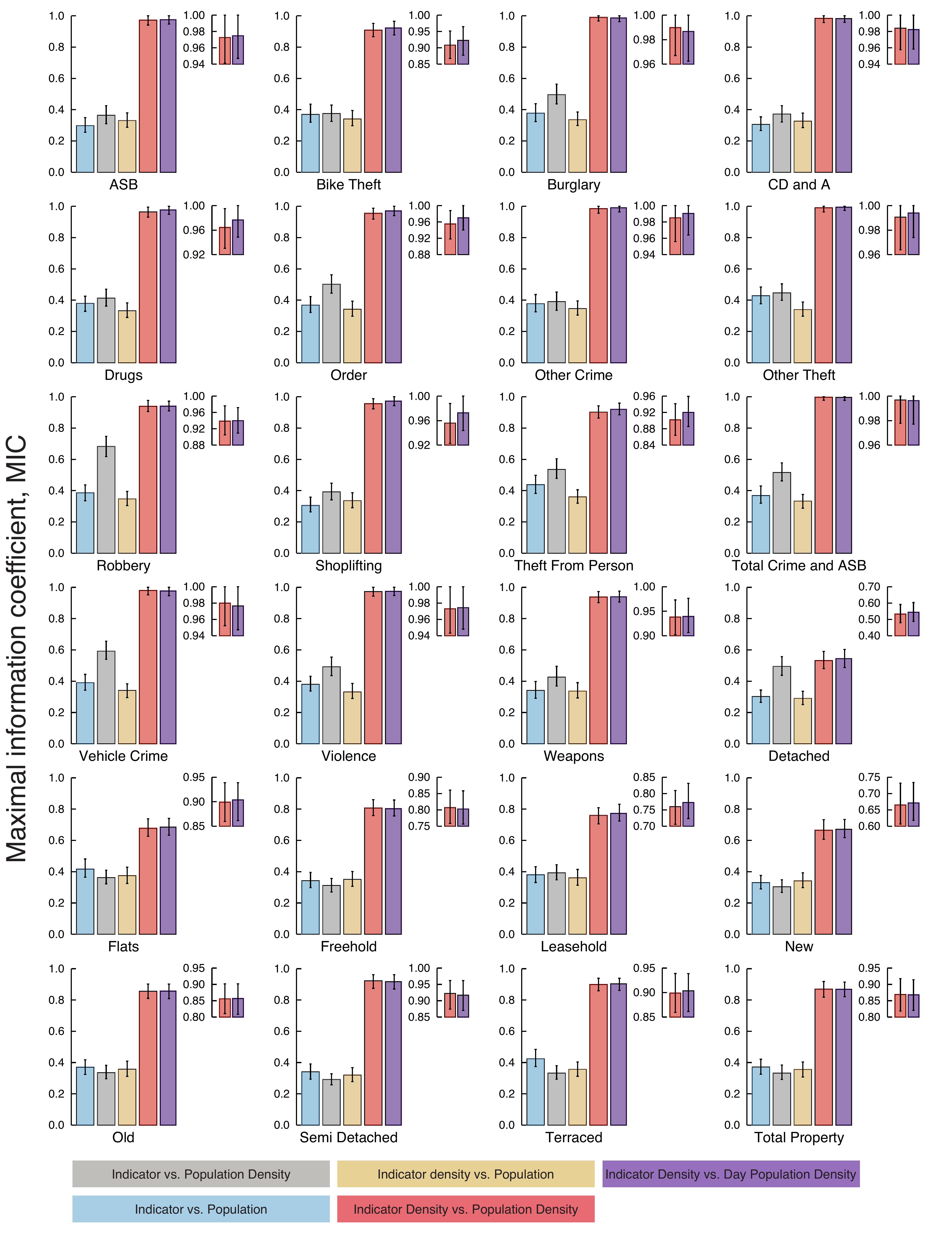}
\end{center}
\textbf{Comparison of  maximal information coefficient (MIC) for different property and crime types.} Similarly to adjusted $R^2$, markedly improved correlations are observed using density metrics which were superior in all cases. Here the error bars stand for 99\% confidence interval obtained via bootstrap. Unlike adjusted $R^2$, MIC indicates no significant difference between population density and day population density (via bootstrap two-sample mean test with 99\% confidence) for other theft and shoplifting.

\subsection*{S2 Fig.}
\label{S2_Fig}
\begin{center}
\includegraphics[scale=0.32]{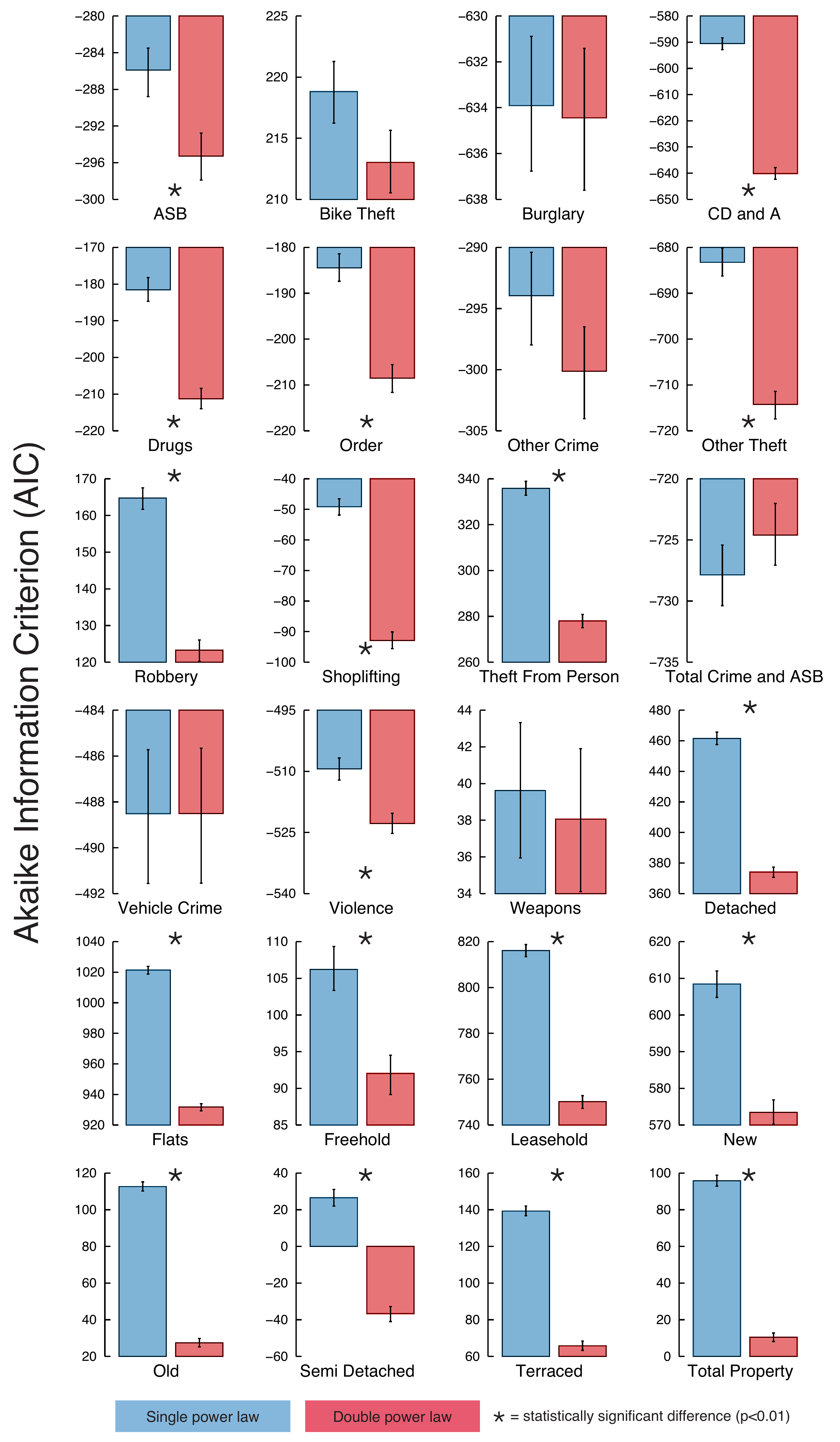}
\end{center}
\textbf{Comparison of the single power-law model (Eq.~\ref{eq_usualscaling_d}) and the double power-law model (Eq.~\ref{eq_doublescaling}) using the Akaike Information Criterion (AIC).} Error bars stand for 99\% bootstrap confidence intervals and the asterisk marks indicate a significant difference (via bootstrap two-sample mean test with 99\% confidence). Notice that the AIC criteria differs from the adjusted $R^2$ only for bike theft.

\subsection*{S3 Fig.}
\label{S3_Fig}
\begin{center}
\includegraphics[scale=0.32]{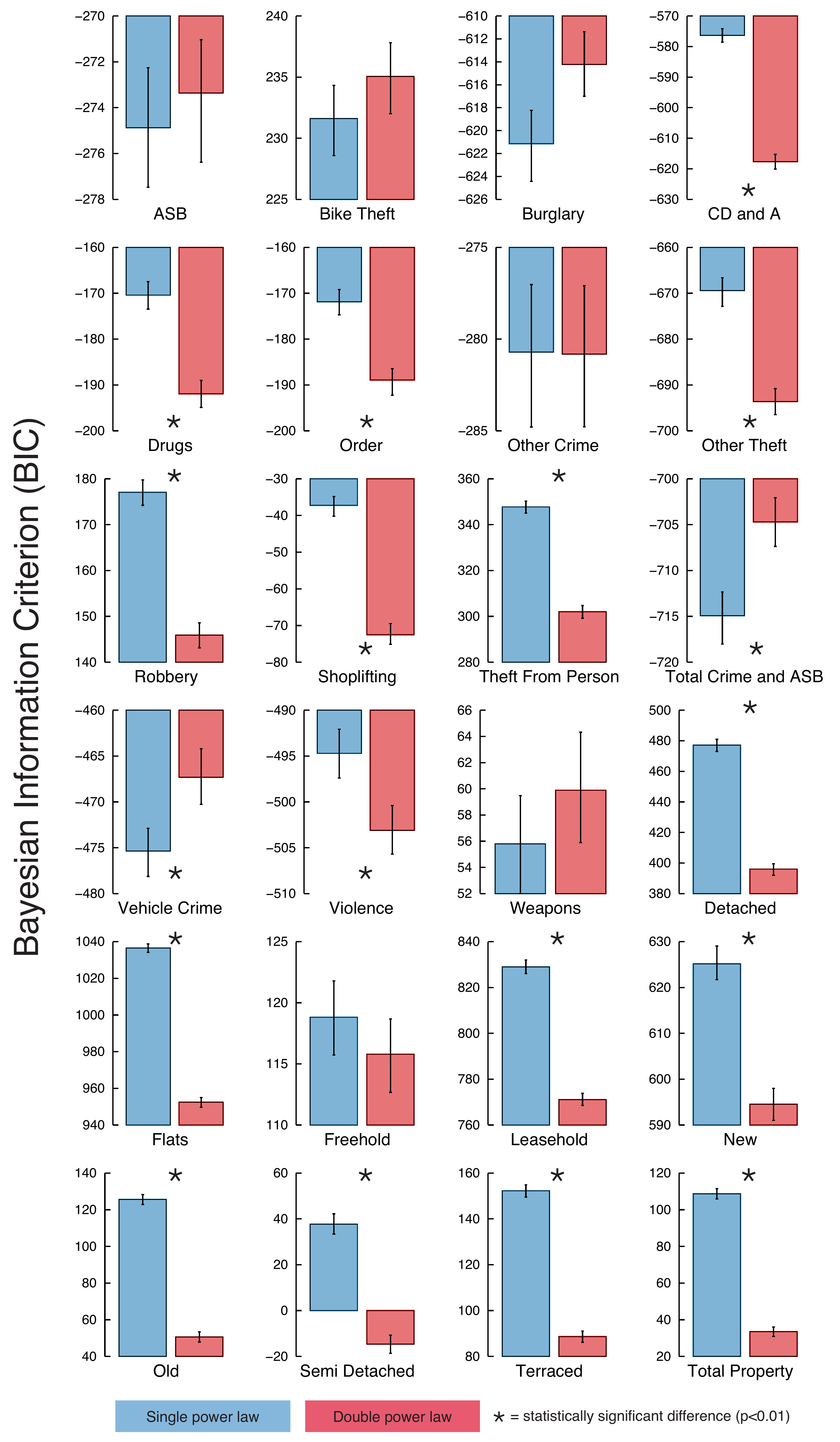}
\end{center}
\textbf{Comparison of  the single power-law model (Eq.~\ref{eq_usualscaling_d}) and the double power-law model (Eq.~\ref{eq_doublescaling}) using the Bayesian Information Criterion (BIC).} Error bars stand for 99\% bootstrap confidence intervals and the asterisk marks indicate a significant difference (via bootstrap two-sample mean test with 99\% confidence). Notice that the BIC criteria differs from the adjusted $R^2$ only for ASB, bike theft and freehold.

\subsection*{S4 Fig.}
\label{S4_Fig}
\begin{center}
\includegraphics[scale=0.4]{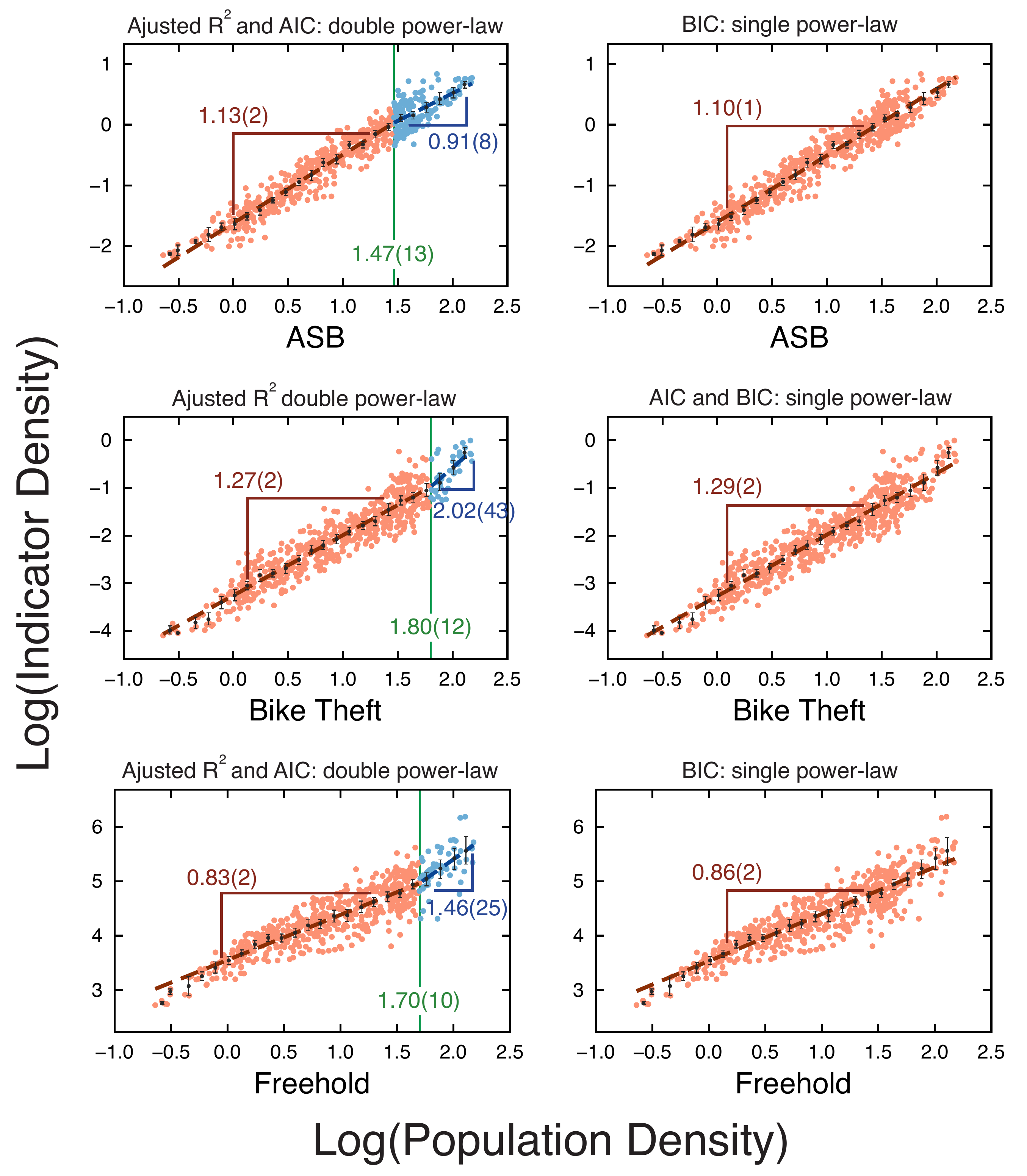}
\end{center}
\textbf{Comparison of  the double power-law model statistics for adjusted $R^2$, AIC and BIC for ASB, Bike Theft and Freehold Property (Eq.~\ref{eq_doublescaling}) } These three metrics were the only cases where the criteria diverged. These can be considered marginal cases of urban scaling transitions.

\section*{Acknowledgements}
The authors are grateful to the Office of National Statistics and the UK Home Office for making these data publicly available. HVR thanks the financial support of CNPq (grant 440650/2014-3)

\section*{Authors Contributions}
Conceived and designed the experiments: QH and HVR. Performed the experiments: QH, HVR, and  DL. Analyzed the data: QH and HVR . Contributed reagents/materials/analysis tools: QH, HVR, and DL . Wrote the paper: QH, HVR, and  DL.

\nolinenumbers

%
%
%

\end{document}